\documentclass[prl,aps,twocolumn,floatfix,longbibliography]{revtex4-2}
\usepackage[utf8]{inputenc}
\usepackage{natbib}
\usepackage{graphicx}
\usepackage{color}
\usepackage[tbtags]{amsmath}
\usepackage{amssymb}
\usepackage{gensymb}
\usepackage{float}

\begin{document}

\title{Current-enabled optical conductivity of superconductors}

\author{Micha{\l} Papaj}
\affiliation{Department of Physics, University of California, Berkeley, CA 94720, USA}
\author{Joel E. Moore}
\affiliation{Department of Physics, University of California, Berkeley, CA 94720, USA}
\affiliation{Materials Sciences Division, Lawrence Berkeley National Laboratory, Berkeley, CA 94720, USA}

\begin{abstract}
In most superconductors, optical excitations require impurity scattering or the presence of multiple bands. This is because in clean single-band superconductors, the combination of particle-hole and inversion symmetries prevents momentum-conserving transitions. In this work we show how the flow of supercurrent can lead to new contributions to optical conductivity. As supercurrent breaks inversion symmetry, transitions across the superconducting gap become allowed even in clean superconductors and dominate over impurity-induced contributions for energies comparable to the gap width. The response is dependent on the nature of the underlying normal state as well as on the type of superconducting order. Through use of the screening supercurrent with controllable magnitude and direction, that arises from an external magnetic field, this enables a detailed investigation of the superconducting state and possible gap symmetry determination in unconventional superconductors for which other techniques have not been practicable. 
\end{abstract}

\maketitle

\textit{Introduction}.---
Optical measurements are one of the most fundamental experimental techniques enabling the studies of quantum materials \cite{orensteinUltrafastSpectroscopyQuantum2012, giannettiUltrafastOpticalSpectroscopy2016}. Properties such as reflectivity and transmissivity can shed light on the electronic structure of solids and enable characterization of the ordered phases in many systems \cite{basovElectrodynamicsCorrelatedElectron2011}. In particular, optical measurements can give insight into the nature of the superconducting state, for example by determination of the superconducting gap size \cite{degiorgiOpticalMeasurementsSuperconducting1994, proninDirectObservationSuperconducting1998, gorshunovOpticalMeasurementsSuperconducting2001}. On the theoretical level, optical properties can be characterized by optical conductivity $\sigma(\omega)$, which can be obtained from microscopic considerations. In the case of superconductors, such a description has been provided by Mattis and Bardeen \cite{mattisTheoryAnomalousSkin1958}, who have analyzed the problem in the dirty limit, where the superconducting coherence length $\xi_0$ is much larger than the mean free path $l$. In this limit the optical response largely follows the normal state Drude conductvity for $\hbar \omega \gg 2\Delta$, where $\Delta$ is the magnitude of the superconducting order parameter. However, the real part of $\sigma(\omega)$ becomes suppressed for smaller frequencies and vanishes for $\hbar \omega \leq 2\Delta$. This theory, together with its extensions to arbitrary purity \cite{leplaeDerivationExpressionConductivity1983, zimmermannOpticalConductivityBCS1991}, has been very successful in explaining the optical properties of many superconductors.

The reasons for the considerable success of Mattis-Bardeen results even beyond the dirty limit have recently been elucidated in a theory for optical transitions of clean, multiband superconductors \cite{ahnTheoryOpticalResponses2021}. Those authors have shown that, due to a combination of inversion and particle-hole symmetries, a selection rule forbids momentum-conserving optical transitions across the superconducting gap in simple single-band superconductors. They have also shown that when multiple bands are present, some transitions become allowed, giving rise to new optical conductivity contributions that can dominate over Mattis-Bardeen terms in very clean systems ($l \gg \xi_0$), such as FeSe. Moreover, it can be shown that when a superconductor breaks inversion symmetry, optical transitions become allowed and the material will exhibit a variety of linear and nonlinear optical effects \cite{xuNonlinearOpticalEffects2019}. While intrinsic inversion-breaking superconductors are rare, another opportunity for breaking inversion opens up when we consider supercurrent flow through the material, as currents are known to strongly affect the optical properties of other materials, such as Dirac and Weyl semimetals \cite{takasanCurrentinducedSecondHarmonic2021}.

In this work, we investigate the effect of inversion-breaking supercurrent on optical conductivity of superconductors. We demonstrate that optical transitions are possible even in clean, single-band superconductors when the flow of supercurrent is introduced. By treating optical conductivity at the linear response level, we show that the predicted signal depends on the nature of the normal state as well as the type of superconducting order. This is corroborated by the comparison of supercurrent-induced responses between single band and Dirac fermion systems, and between s-wave and d-wave pairings. The predicted optical response dominates over that of Mattis-Bardeen theory for photon energies in the vicinity of the superconducting gap edge. As the supercurrent flow can be established and controlled by applying external magnetic field through the Meissner effect, this approach introduces a control knob that can modify an optical response of a superconductor in an experimental setting without requiring the system to be driven far from equilibrium. Combining these factors leads to a promising tool for investigation of the superconducting state.

\begin{figure}
    \centering
    \includegraphics[width=0.99\columnwidth]{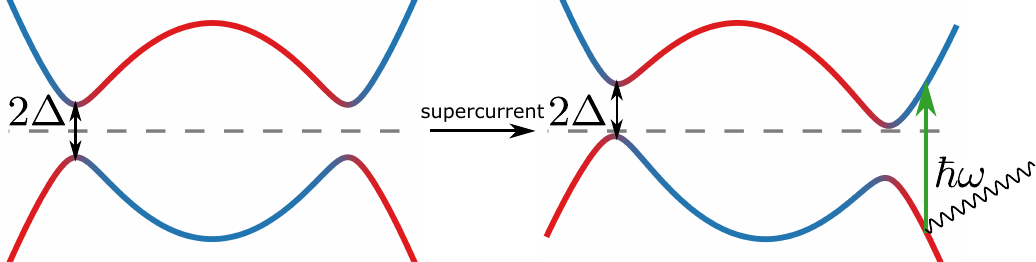}
    \caption{The effect of supercurrent on quasiparticle dispersion. In the absence of supercurrent, transitions across the superconducting gap are forbidden. However, inversion breaking due to the current flow enables transitions that contribute to interband optical conductivity. Blue and red colors indicate degree of superposition between particle and hole-like states.}
    \label{fig:dispersion}
\end{figure}

\textit{Supercurrent and the excitation spectrum}.---
When a supercurrent flow is introduced in a superconductor, the Cooper pairs in the condensate acquire finite momentum $2\mathbf{q}$. As a result, the dispersion of quasiparticle excitations now includes a term corresponding to a Doppler shift \cite{fuldeGaplessSuperconductingTunnelingTheory1969}:
\begin{equation}
\label{eq:doppler_effect}
    E(\mathbf{k}) = \sqrt{\xi_\mathbf{k}^2 + \Delta^2} + v_\mathbf{k} \cdot \mathbf{q},
\end{equation}
with $\xi_\mathbf{k} = \epsilon_\mathbf{k} - \mu$, where $\epsilon_\mathbf{k}$ is the particle dispersion in the normal state, $\mu$ is the chemical potential, $\Delta$ is the superconducting order parameter, and $v_\mathbf{k}=\partial \epsilon_\mathbf{k}/\partial \mathbf{k}$ is the group velocity. As the Cooper pair momentum $2\mathbf{q}$ is determined by the direction of the supercurrent, the quasiparticle energy increases or decreases, depending whether it moves parallel or anti-parallel to the current. Therefore, supercurrent flow introduces anisotropy into the quasiparticle dispersion, in simple cases leading to tilting of the spectrum. This is presented in Fig.~\ref{fig:dispersion}, which shows the Bogoliubov-de Gennes (BdG) spectrum around the superconducting gap. When the Cooper pair momentum exceeds a critical value, Eq.~\eqref{eq:doppler_effect} allows for zero energy excitations. This leads to the appearance of segmented Fermi surface, which has recently been observed in thin films of 3D topological insulators under proximity effect \cite{zhuDiscoverySegmentedFermi2021} and can lead to topological phase transition \cite{papajCreatingMajoranaModes2021, takasanSupercurrentinducedTopologicalPhase2021}. A related phenomenon in which the Doppler effect plays a role is the Volovik effect \cite{volovikSuperconductivityLinesGap1993}, where the supercurrent in vortices leads to changes in the density of states. This effect has been discussed and detected in optical measurements previously \cite{mallozziHighFrequencyElectrodynamicsMathrmBi1998, kimMicrowaveConductivityHighpurity2004, tagayBCSWaveBehavior2021}. However, in our case we are concerned with small in-plane magnetic fields that do not lead to formation of vortices. Supercurrent was also explored in the context of infrared activation of the Higgs mode in superconductors \cite{moorAmplitudeHiggsMode2017, nakamuraInfraredActivationHiggs2019}.

The supercurrent flow can arise either due to an explicit transport current or due to an external applied magnetic field \cite{anthoreDensityStatesSuperconductor2003}. In the latter case, as a result of the Meissner effect, a screening supercurrent develops at the surface of the superconductor. This screening supercurrent is directly connected to the magnetic vector potential $\mathbf{A}$ via the London equation:
\begin{equation}
    \mathbf{j}_S = - \frac{n_S e^2}{m} \mathbf{A},
\end{equation}
where $n_S$ is the superfluid density, $e$ is the electron charge and $m$ is the electron mass. The behavior of $\mathbf{A}$ at the surface of a superconductor can be determined by combining London and Maxwell equations. Assuming that the boundary of the superconductor is at $z=0$ plane and the supercurrent flows along $x$ direction, in London gauge ($\nabla \cdot \mathbf{A} = 0$) the vector potential will only have a non-zero $x$ component. At the surface the vector potential can thus be determined to be $A_x(z=0) = B_\text{ext} \lambda_L$, where $B_{ext}$ is the magnitude of the external magnetic field, which is pointing along $y$ direction ($\mathbf{B}(z>0) = B_\text{ext} \hat{y}$), and $\lambda_L$ is the London penetration depth. Therefore, in order to obtain a larger Cooper pair momentum $2q = 2 e B_\text{ext} \lambda_L$ due to the external magnetic field, one should increase the external magnetic field and use superconductors with longer $\lambda_L$. With Cooper pair momentum and supercurrent present, the inversion symmetry is broken and we can now investigate the new contributions to optical conductivity.

\textit{Superconductor models}.---
In this work we focus on two different models of superconductors that exemplify the different aspects of supercurrent-enabled optical conductivity. To study these types of superconductors, we employ BdG formalism to calculate the Matsubara Green's functions as discussed below. We obtain results for single-band spin degenerate s-wave superconductor and Dirac fermion under proximity effect from s-wave superconductor. The results can also be extended to d-wave superconductors as shown in Supplemental Materials \cite{SM}. The mean-field Hamiltonian is assumed to arise from interacting Hamiltonian $H=\sum_{\mathbf{k}\sigma} \xi_\mathbf{k} c^\dagger_{\mathbf{k}\sigma} c_{\mathbf{k}\sigma} + \sum \lambda_\mathbf{q} c^\dagger_{\mathbf{k}+\mathbf{q}\sigma} c^\dagger_{\mathbf{k}'-\mathbf{q}\sigma'}c_{\mathbf{k}'\sigma'}c_{\mathbf{k}\sigma}$, leading to a gap equation at zero temperature:
\begin{equation}
    \Delta_\mathbf{k} = - \sum_\mathbf{p} \lambda_{\mathbf{k}-\mathbf{p}} \frac{\Delta_\mathbf{p}}{2\sqrt{\xi_\mathbf{p}^2 + \Delta_\mathbf{p}^2}}
\end{equation}
In the s-wave case, the interaction strength is momentum independent $\lambda_\mathbf{p} = \lambda$ and in consequence, the superconducting order parameter is also momentum independent $\Delta_\mathbf{k} = \Delta$.

In each case, the influence of the supercurrent is introduced by including the vector potential using minimal coupling $k_i \rightarrow k_i + q_i \tau_z$, where $\tau_z = \pm 1$ for particle and hole sector of mean-field BdG Hamiltonian. In all of the following calculations we will assume that the supercurrent, and thus the Cooper pair momentum, is directed along $x$ axis and so $\mathbf{q} = q_x \hat{x}$. In the case of analytical solution for the optical conductivity of proximitized Dirac fermion, we do not solve for the superconducting order parameter self-consistently when supercurrent is present. Still, for small supercurrent this should not introduce qualitative differences \cite{horvathCurrentcarryingStateNodal2012}, which we also verify numerically in the tight-binding model case.

The most generic model of superconductor that we consider consists of a single spin-degenerate tight-binding band with nearest neighbor hopping $t$ on a square lattice with unit lattice constant at chemical potential $\mu$ with a superconducting gap $\Delta$:
\begin{equation}
\label{eq:tight_binding_ham}
    H_\text{BdG}^\text{TB}(\mathbf{k}) = \left(t(2 - \cos(k_x) - \cos(k_y)) - \mu \right) \tau_z + \Delta \tau_x
\end{equation}
Such a simple model can nevertheless fully demonstrate the supercurrent-induced optical conductivity.

For the purpose of analytical derivation we also consider a Dirac fermion with s-wave superconducting order parameter:
\begin{equation}
    H_\text{BdG}^\text{D}(\mathbf{k}) = \left(\hbar v k_x s_y - \hbar v k_y s_x - \mu \right) \tau_z + \Delta \tau_x
\end{equation}
where $s_i$ are Pauli matrices representing the spin degree of freedom. This model can describe the surface state of a 3D topological insulator under proximity effect from a conventional superconductor, as in the case of the recent experiment reporting observation of a segmented Fermi surface \cite{zhuDiscoverySegmentedFermi2021}. In such a scenario, the Fermi energy is placed high above the Dirac point and so for small photon energies we can focus only on the upper Dirac cone around the superconducting gap. This allows us to treat this system as effectively a single-band superconductor with a helical spin-texture \cite{SM}.

\begin{figure}
    \centering
    \includegraphics[width=0.99\columnwidth]{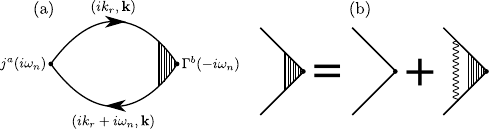}
    \caption{Feynman diagrams in Matsubara formalism. (a) Current-current correlation function evaluated using a bubble diagram with vertex correction to ensure satisfaction of Ward identities. (b) Self-consistent equation for the vertex correction. Straight lines indicate propagators in Matsubara formalism and wiggly line indicates interaction that leads to superconducting pairing.}
    \label{fig:feynman_diagrams}
\end{figure}

\textit{Formalism}.---
To obtain the optical conductivity $\sigma_{ab}(\omega)$ we work within linear response using the Kubo formula in Matsubara formalism adapted to the BdG approach. This corresponds to evaluation of the current-current correlation function as depicted in Fig.~\ref{fig:feynman_diagrams}(a). Expressed in terms of Matsubara Green's functions this gives \cite{colemanIntroductionManyBodyPhysics2015}:
\begin{align}
    \Pi_{ab}&(i\omega_n) = \notag \\ & -\frac{1}{\beta V} \sum_{\mathbf{k}, ik_r} \mathrm{Tr} \, j_0^a G_0(\mathbf{k}, ik_r + i\omega_n) \Gamma^b(i\omega_n) G_0(\mathbf{k}, ik_r),
\end{align}
where $j_0^a$ is the component of bare current operator for the BdG Hamiltonian as described below, $\Gamma_b$ is the current operator with vertex correction included, and $G_0(\mathbf{k}, ik_n) = (ik_n - H_\mathrm{BdG})^{-1}$ is the Matsubara Green's function for the BdG Hamiltonian. From this we can obtain the real (dissipative) part of the optical conductivity by analytic continuation:

\begin{equation}
   \mathrm{Re}\, \sigma_{ab}(\omega) = \frac{1}{\omega} \mathrm{Im} \Pi_{ab}(\omega + i \eta)
\end{equation}

The vertex correction $\Gamma$ is crucial in consideration of optical responses in the presence of the supercurrent as it ensures the satisfaction of Ward identities and in turn that the obtained results are physical. Indeed, calculating the uncorrected current-current correlation function for the case of supercurrent flow in a system with a parabolic band yields a non-zero result \cite{SM}. However, since the system with a parabolic band is Galilean invariant, the current operator is proportional to momentum $\mathbf{j}_0\sim\mathbf{k}$, which means it commutes with the full interacting Hamiltonian $[j^a_0, H] = 0$. Therefore, no non-trivial optical response is possible. In order to rectify this issue, we include the vertex correction at the ladder approximation level, which constitutes an appropriate conserving approximation for the superconductivity treated at the mean-field level \cite{daiOpticalConductivityPair2017, nambuQuasiParticlesGaugeInvariance1960}. Such a vertex correction is depicted in Fig.~\ref{fig:feynman_diagrams}(b) and corresponds to the following self-consistent equation:
\begin{align}
    \Gamma^a&(i\omega_n) = \notag \\& j_0^a -\frac{\lambda}{\beta V} \sum_{\mathbf{k}, ik_r} \tau_z G_0(\mathbf{k}, ik_r+i\omega_n)\Gamma^a(i\omega_n) G_0(\mathbf{k}, ik_r) \tau_z
\end{align}
For the s-wave case, since the interaction strength is momentum-independent, the correction to the vertex is only a function of frequency, constant in momentum space.

The bare current operators in each of the models are calculated including an infinitesimal vector potential perturbation $\delta \mathbf{A}$ in the normal state Hamiltonians in particle-hole space according to minimal coupling rule $\mathbf{k} \rightarrow \mathbf{k} - e \delta \mathbf{A}\tau_z$, where $\tau_z = \pm 1$ for particle and hole sectors of BdG Hamiltonian. By taking appropriate derivative we arrive at the current operator in a given direction:
\begin{equation}
    j_0^a = -\frac{\partial H_\text{BdG}(\mathbf{k} - e \delta \mathbf{A}\tau_z)}{\partial \delta A_a} \Bigg|_{\delta\mathbf{A}=0, \Delta=0}
\end{equation}

\begin{figure}
    \centering
    \includegraphics[width=0.99\columnwidth]{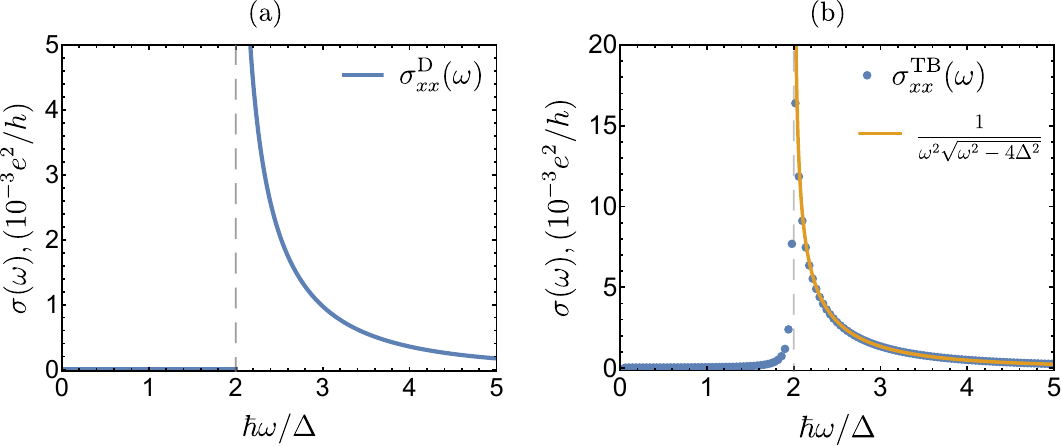}
    \caption{Real part of optical conductivity in a superconductor carrying supercurrent. (a) Analytical expression for $\sigma_{aa}(\omega)$ for s-wave superconductor with Dirac fermion dispersion. As s-wave superconductors are fully gapped for small Cooper pair momentum, the optical conductivity only carries the interband contribution and remains gapped for $\hbar \omega < 2 \Delta$. (b) Numerical results for $\sigma_{xx}(\omega)$ calculated using tight-binding model. The solid line shows $1/(\omega^2\sqrt{\omega^2-4\Delta^2})$ dependence that can be inferred from analytical calculation without vertex correction.}
    \label{fig:sigma}
\end{figure}
\textit{Optical conductivity results}.---
We can now employ the formalism described above to obtain the real part of the optical conductivity. Since the s-wave superconductors have no Fermi surface for small Cooper pair momenta, only the interband contribution is relevant to their optical conductivity at $T=0$. We begin with the proximitized Dirac fermion as the analytical result allows for gaining better insight into the phenomenon. In this case we obtain the result including the vertex correction \cite{SM}:
\begin{equation}
    \text{Re}\, \sigma_{aa}^{D}(\omega) = \frac{e^2}{h} \frac{ \pi}{4} \frac{\Delta}{\mu} \frac{\hbar^2 v^2 q_x^2 \Delta}{\hbar^2 \omega^2 \sqrt{\hbar^2 \omega^2 - 4 \Delta^2}} \Theta(\omega - 2 \Delta),
\end{equation}
with $a=x,y$, which means optical conductivity for Dirac fermion case is equal in the directions parallel and perpendicular to the supercurrent. This expression, presented in Fig.~\ref{fig:sigma}(a), highlights several important characteristics of supercurrent-induced optical conductivity. First of all, since the supercurrent in this case only tilts the dispersion of quasiparticles, the energy separation of the two BdG branches of the spectrum at a given momentum $\mathbf{k}$ remains the same. Therefore, the minimal photon energy at which a transition can occur is still $2\Delta$ and optical conductivity remains zero for smaller energies. Moreover, the optical conductivity also has a $1/\sqrt{\hbar \omega - 2\Delta}$ singularity at the gap boundary, which will correspond to a peak in experimentally relevant scenarios. The origin of this singularity is related to the density of states of a BCS superconductor, which contains precisely this type of singularity at the gap edges. Finally, $\text{Re}\, \sigma_{aa}^{D}(\omega)$ depends quadratically on the Cooper pair momentum $\mathbf{q}$.

In the case of tight-binding band described by Eq.~\eqref{eq:tight_binding_ham} we cannot obtain an analytical result and we have to rely on numerical calculation for the current-current correlation function with vertex correction as shown in Fig.~\ref{fig:sigma}(b). In numerical calculations we have used $t=1$, $\mu=0.9$, $\lambda=-1$, and $q_x=0.025$. As mentioned above, approximating the band as parabolic would lead to no non-trivial optical response due to Galilean invariance. Nevertheless, calculation using bare current vertices can give additional insight into the functional form of frequency dependence, leaving the exact prefactor (dependent on the deviation from parabolic dispersion) to be determined numerically. As discussed in Supplemental Materials, the frequency dependence is $\sim 1/(\omega^2\sqrt{\omega^2-4\Delta^2})$ and can be precisely fitted to the numerical results as demonstrated in Fig.~\ref{fig:sigma}(b). This means that the results for tight-binding band shares similarities with the Dirac fermion case, stemming from the same s-wave type of order parameter. Those features are the presence of the singularity at the gap edge and absence of optical absorption inside of the gap. However, in contrast to the Dirac fermion, in this case the optical conductivity in the direction perpendicular to the supercurrent ($\sigma_{yy}$) vanishes. This signifies the impact of the normal state dispersion on the detailed characteristics of supercurrent-induced optical conductivity.

\begin{figure}
    \centering
    \includegraphics[width=0.75\columnwidth]{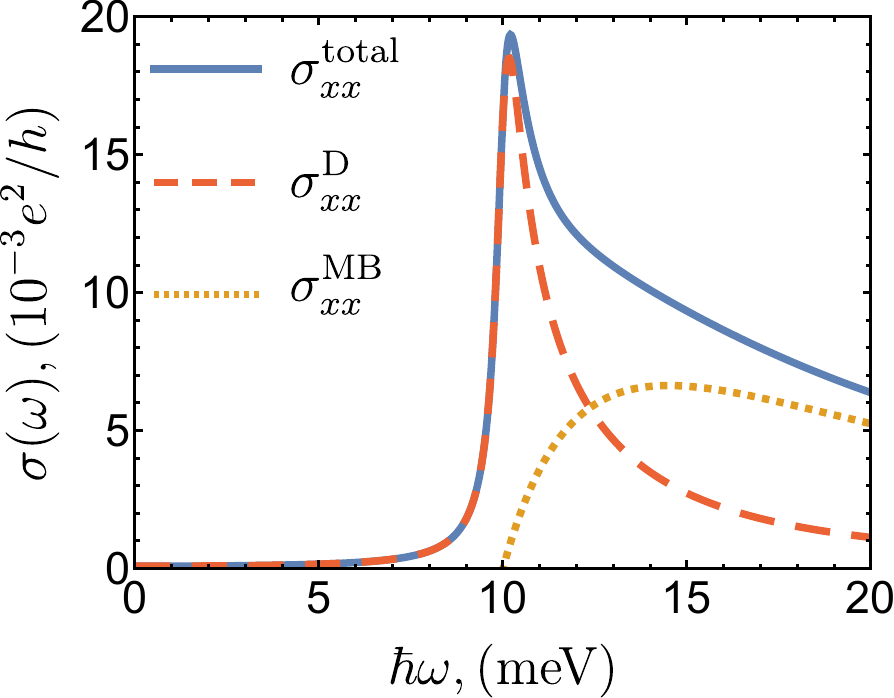}
    \caption{Optical conductivity for superconducting Dirac surface state of CaKFe$_4$As$_4$ at $T = 1\,\text{K}$, with $\Delta = 5\,\text{mev}$, $v=10^5 \text{m/s}$ and $\mu = 20\,\text{meV}$ with $l/\xi_0 = 8$, for supercurrent that closes half of the superconducting gap. Yellow dotted line shows corresponding Mattis-Bardeen contribution from impurity scattering, red dashed line shows supercurrent-induced part, and blue solid line shows both contributions combined. }
    \label{fig:sigma_numerical}
\end{figure}
\textit{Discussion}.---
As presented above, the characteristics of supercurrent-induced optical conductivity vary considerably depending both on the nature of the underlying normal state as well as the type of the superconducting order parameter. As a consequence, optical conductivity with supercurrent present may serve as an important tool in characterization of the superconducting state. In particular, as the supercurrent flow can be introduced by applying external magnetic field and utilizing the Meissner effect, the anisotropies of both the normal state dispersion as well as the superconducting gap can be investigated using a vector magnet. As was shown in the Bi$_2$Te$_3$/NbSe$_2$ system \cite{zhuDiscoverySegmentedFermi2021}, applying only 20 mT of in-plane field was sufficient to realize Doppler energy shift comparable to the superconducting gap. At the same time, that experiment has demonstrated that the screening supercurrent effects are sensitive to magnetic field direction. Therefore, similar effects may be visible in optical conductivity measurements. This could enable disentangling the various Fermi pockets that contribute to superconductivity in materials such as some iron-based superconductors, where superconductivity in bulk bands and Dirac surface states coincides. This method could also elucidate the nature of the superconducting order parameter in recently studied moir\'e superconductors such as twisted bilayer and trilayer graphene \cite{caoUnconventionalSuperconductivityMagicangle2018, caoPaulilimitViolationReentrant2021}.

To estimate the visibility of the proposed effect, it is necessary to evaluate it in comparison to the Mattis-Bardeen optical conductivity $\sigma^{MB}$ that arises purely from impurity effects. As an example, we compare the supercurrent-induced and impurity-driven optical conductivity for clean iron superconductors with Dirac surface states (such as CaKFe$_4$As$_4$ \cite{liuNewMajoranaPlatform2020}). The results of such comparison are presented in Fig.~\ref{fig:sigma_numerical}. The supercurrent-induced effect dominates over $\sigma^{MB}$ in the vicinity of the superconducting gap. This is the region, where the difference between the two sources of interband transitions is the most apparent:  while $\sigma^{MB}$ follows Drude tail for large $\omega$, it gets suppressed at the gap edge. In contrast, supercurrent flow introduces singular behavior of $\sigma(\omega)$ at the gap edge, leading to the appearance of a sharp peak. Moreover, in real materials the optical conductivity will depend on the direction of the supercurrent with respect to crystalline axes due to corrections to dispersion such as hexagonal warping \cite{fuHexagonalWarpingEffects2009}. Since both the magnitude of the peak as well as the anisotropy of optical conductivity can be controlled by the direction of the current (and in turn, the direction of the external magnetic field), it is possible to clearly distinguish current-enabled and impurity effects.

In summary, we have shown that introducing current flow in a superconductor, either through applying external magnetic field or by direct transport current, can significantly affect its optical properties at photon energies close to the superconducting gap magnitude. The effect is generic, appearing independently of the nature of the normal states as well as the pairing symmetry, yet it is sensitive to both of these important material characteristics. As such, supercurrent-driven optical conductivity may become a valuable tool in investigations of novel superconductors.

\textit{Note added:} We thank Liang Fu and Philip Crowley for bringing our attention to the issue of Galilean invariance and its impact on the optical response. They have recently studied the optical response of s-wave superconductors with supercurrent using a different formalism than ours, and our results are in agreement in that case \cite{crowleySupercurrentInducedResonant2022}.

\begin{acknowledgements}
 The authors also acknowledge helpful discussions with J. Orenstein. This work was supported by the Quantum Science Center (QSC), a National Quantum Information Science Research Center of the U.S. Department of Energy (DOE).  M.P. received additional fellowship support from the Emergent Phenomena in Quantum Systems program of the Gordon and Betty Moore Foundation and J.E.M. acknowledges a Simons investigatorship.
\end{acknowledgements}

\bibliography{optical_conductivity_SC}

\onecolumngrid

\newpage

\begin{center}
\textbf{\large Supplemental Materials for "Current-enabled optical conductivity of superconductors"}
\end{center}
\setcounter{equation}{0}
\setcounter{figure}{0}
\setcounter{page}{1}
\renewcommand{\thefigure}{S\arabic{figure}}
\renewcommand{\theequation}{S\arabic{equation}}

\section{Projection to superconductivity in upper Dirac cone}
In this section we discuss how for large chemical potential $\mu$ we treat Dirac fermion under proximity effect as a single-band superconductor. The BdG Hamiltonian for Dirac fermion under proximity effect and with supercurrent flowing is:

\begin{equation}
    H(\mathbf{k}) = \begin{pmatrix}
     -\mu  & - \hbar v (k_y+i (k_x+q_x)) & \Delta  & 0 \\
  \hbar v (-k_y+i (k_x+q_x)) & -\mu  & 0 & \Delta  \\
 \Delta  & 0 & \mu  &  \hbar v (k_y+i (k_x-q_x)) \\
 0 & \Delta  &  \hbar v (k_y-i (k_x-q_x)) & \mu  \\
    \end{pmatrix}
    \label{eq:S_BdG_Dirac}
\end{equation}
In absence of superconducting pairing ($\Delta = 0$), its eigenvalues are given by $E_{s_1,s_2}(\mathbf{k}) = s_1 \hbar v \sqrt{(k_x + s_2 q_x)^2 + k_y^2} - s_2 \mu$, where $s_1 = \pm 1$ and $s_2 = \pm 1$. These eigenvalues have corresponding eigenvectors $\psi_{s_1, s_2}$:
\begin{equation}
    \psi_{s_1, -1} = \left(0, 0, s_1 i \exp\left( -i \alpha_-\right), 1 \right)^T/\sqrt{2}, \quad \quad
    \psi_{s_1, +1} = \left(-s_1 i \exp\left( -i \alpha_+\right), 1, 0, 0 \right)^T/\sqrt{2}
    \label{eq:S_basis}
\end{equation}
where $\alpha_\pm = \arctan(k_x \pm q_x, k_y)$. Since we are interested in the properties of the system at low energies, we can focus on only a single band and its BdG partner, which are given by $s_1=1$, $s_2 = 1$, and $s_1=-1$, $s_2 = -1$. We can now transform the Hamiltonian of Eq.~\eqref{eq:S_BdG_Dirac} to the basis given by Eq.~\eqref{eq:S_basis}. When chemical potential is large compared to all the other energy scales of the problem, we can neglect the coupling to the bands that are further away from Fermi energy. When we expand the transformed Hamiltonian to the lowest order in $q_x$, we obtain:

\begin{equation}
    H_\text{projected}(\mathbf{k}) = \begin{pmatrix}
     \hbar v k-\mu +\hbar v q_x \frac{k_x}{k} & \Delta  \\
 \Delta  & - \hbar v k+\mu +\hbar v q_x\frac{k_x}{k} \\
    \end{pmatrix}
\end{equation}
In this approximation, for large chemical potential Dirac fermion under proximity effect from an s-wave superconductor can be effectively described as a single band superconductor. To apply this in the optical conductivity calculation in the main text, we transform other operators to this basis as well.

\section{Optical conductivity of superconductors due to impurities}
In the main text we compare the supercurrent-induced optical conductivity to the effects resulting from impurity scattering. For s-wave superconductors we use the original results of Mattis and Bardeen \cite{mattisTheoryAnomalousSkin1958}, which are given in terms of the normal state Drude conductivity $\sigma_N(\omega)$:

\begin{equation}
    \frac{\text{Re}\,\sigma^{MB}_\text{s-wave}(\omega)}{\text{Re}\, \sigma_N(\omega)} = \frac{(2 \Delta +\hbar \omega )}{\hbar \omega} E\left(\frac{(\hbar\omega -2 \Delta )^2}{(\hbar\omega + 2 \Delta )^2}\right)-\frac{4 \Delta}{\hbar\omega}  K\left(\frac{(\hbar\omega -2 \Delta )^2}{(\hbar\omega + 2 \Delta)^2}\right), \quad \quad \hbar \omega > 2 \Delta
\end{equation}
where $E(z)$ and $K(z)$ are complete elliptic integrals of the first and second kind, respectively. This form is responsible for suppression of the real part of optical conductivity at the superconducting gap edge. For large photon frequencies, $\text{Re}\,\sigma^{MB}(\omega)/\text{Re}\, \sigma_N(\omega) \approx 1$ and the optical conductivity of a superconductor follows what remains of the Drude conductivity tail.

In the case of d-wave superconductor, we use the following approximate expression \cite{yanagisawaChapterOpticalProperties2007}:
\begin{align}
    &\frac{\text{Re}\,\sigma^{MB}_\text{d-wave}(\omega)}{\text{Re}\, \sigma_N(\omega)} = \notag \\ &\frac{1}{2 \hbar \omega} \int_{-\infty}^\infty d\epsilon \left( \tanh \left( \frac{\epsilon + \hbar \omega}{2 k_B T} \right) - \tanh \left( \frac{\epsilon}{2 k_B T} \right) \right) \Bigg\langle \text{Re} \frac{1}{\sqrt{1 - \left(\frac{\Delta}{\epsilon + \hbar \omega}\right)^2 \cos^2(2 \theta)}} \Bigg\rangle_\theta \Bigg\langle \text{Re} \frac{1}{\sqrt{1 - \left(\frac{\Delta}{\epsilon}\right)^2 \cos^2(2 \theta)}} \Bigg\rangle_\theta
\end{align}
where the averaging over angle $\theta$ in momentum space is defined as:
\begin{equation}
    \Bigg \langle \text{Re} \frac{1}{\sqrt{1 - t^2 \cos^2(2 \theta)}} \Bigg \rangle_\theta = \text{Re} \frac{1}{2\pi} \int_0^{2\pi} d\theta \frac{1}{\sqrt{1 - t^2 \cos^2(2 \theta)}} = \text{Re} \frac{2}{\pi} K\left( t^2 \right)
\end{equation}
Similarly to the s-wave case, this form reproduces Drude conductivity tail for large frequencies. However, inside of the superconducting gap it does not immediately go to zero due to the nodal character of the gap. Still, the optical conductivity inside of the gap becomes suppressed compared to normal state Drude peak.

\section{Analytical solution for the vertex correction for s-wave superconductors}

For analytical solution of the vertex correction equation we largely follow the procedure outlined in the Appendix of Ref. \cite{daiOpticalConductivityPair2017}. This procedure will work for the single band tight-binding model as well as for the Dirac fermion. This is because, as shown in the first section, for large chemical potential we can project the proximitized Dirac fermion Hamiltonian to just the upper cone, obtaining an effective single band Hamiltonian. We can therefore work with a generic Hamiltonian of the form:

\begin{equation}
    H = \begin{pmatrix} 
    \xi_1(\mathbf{k}) & \Delta \\
    \Delta & \xi_2(\mathbf{k})
    \end{pmatrix}
\end{equation}

Since the interaction strength in the vertex correction equation is momentum-independent, the vertex correction added on top of the bare current operator will also be constant in momentum space and we can express it as coefficients in Pauli matrices expansion as:
\begin{equation}
    \Gamma^a(\mathbf{k}, \omega) = \sum_{i=0}^3 j_0^{a,i}(\mathbf{k}) \tau_i + \sum_{i=0}^3 \Gamma^{a,i} \tau_i
\end{equation}
For the models under consideration in this work $j_0^{a,1}=j_0^{a,2}=0$. Using the generic Hamiltonian to calculate the Matsubara Green's functions and plugging this vertex correction expansion into the self-consistent vertex correction equation we obtain:

\begin{equation}
    \begin{pmatrix}
    \Gamma^{a,1}\\
    \Gamma^{a,2}\\
    \Gamma^{a,3}
    \end{pmatrix}  = \int \frac{d^2k}{(2 \pi)^2} \frac{\lambda}{D(\mathbf{k})} \left(
    \begin{pmatrix}
     - (\xi_1-\xi_2)^2 & i   \omega  (\xi_1-\xi_2) & 2 \Delta    (\xi_1-\xi_2) \\
 -i   \omega  (\xi_1-\xi_2) & -  \left(4 \Delta ^2+(\xi_1-\xi_2)^2\right) & 2 i \Delta    \omega  \\
 2 \Delta    (\xi_2-\xi_1) & 2 i \Delta    \omega  & 4 \Delta ^2   \\
    \end{pmatrix}
    \begin{pmatrix}
    \Gamma^{a,1}\\
    \Gamma^{a,2}\\
    \Gamma^{a,3}
    \end{pmatrix}
    +
    \begin{pmatrix}
    2 \Delta  j_0^{a, 3}   (\xi_1-\xi_2) \\
    2 i \Delta  j_0^{a, 3}   \omega \\
    4 \Delta ^2 j_0^{a, 3} 
    \end{pmatrix}
    \right)
\end{equation}
where $D(\mathbf{k}) = \sqrt{4 \Delta^2 + (\xi_1-\xi_2)^2}\left(4 \Delta^2 + (\xi_1-\xi_2)^2 + (\eta - i\omega)^2\right)$. Due to the form of $D(\mathbf{k})$, for small superconducting gap and frequencies of the same order the main contributions to the integral will come from region of momentum space where $\xi_1 - \xi_2 \approx 0$ and so $\Gamma^{a,1}\approx0$. We can then solve for the remaining $\Gamma^{a,i}$ components and by using the self-consistent gap equation we also find that $\Gamma^{a,3}=0$. We finally obtain:

\begin{equation}
    \Gamma^{a,2}(\omega) = \frac{2i\Delta}{\omega} \frac{I[j_0^{a, 3}]}{I[1]}, \quad \quad I[f(\mathbf{k})] = \int \frac{d^2k}{(2 \pi)^2} \frac{f(\mathbf{k})}{D(\mathbf{k})}
\end{equation}

We can now use this to evaluate the corrected current-current correlation function to find optical conductivity. From the diagram presented in the main text using the fact that $\Gamma^{a,1} = \Gamma^{a,3} = 0$ we obtain:

\begin{equation}
    \Pi_{aa}(\omega) = \int \frac{d^2k}{(2 \pi)^2} \frac{4 \Delta  j_0^{a, 3} (2 \Delta  j_0^{a, 3}+i \Gamma^{a,2} \omega )}{D(\mathbf{k})}
\end{equation}
Plugging in the result for $\Gamma^{a,2}$ we get:
\begin{equation}
    \Pi_{aa}(\omega) = \int \frac{d^2k}{(2 \pi)^2} \frac{8 \Delta^2  j_0^{a, 3} \left(j_0^{a, 3}-\frac{I[j_0^{a, 3}]}{I[1]}\right)}{D(\mathbf{k})}
\end{equation}
Now we can see that when $j_0^{a, 3}$ is momentum-independent, $I[j_0^{a, 3}] = j_0^{a, 3} I[1]$ and the optical conductivity vanishes. This is precisely the case for parabolic system with supercurrent flow, where $j_0^{a, 3} \sim q_x$. This result guarantees that in Galilean invariant system the optical response will vanish as expected. The above equation can also be used to determine analytically optical conductivity for the case of proximitized Dirac fermion, with the result presented in the main text. We have also verified the conclusions above for parabolic system numerically without the approximations taken during the derivation, as shown in the next section.

\section{Supercurrent-induced optical conductivity and Galilean-invariant systems}

In this section we expand upon the discussion of the supercurrent-induced effects in Galilean-invariant systems. As mentioned in the main text, when a single band system has a parabolic dispersion in the normal state, the bare current operator is proportional to momentum ($\mathbf{j}_0 \sim \mathbf{k}$) and since the interactions conserve momentum, $\mathbf{j}_0$ commutes with both the kinetic and the interaction terms in the Hamiltonian. As the Kubo formula relates conductivity to $\langle[j_0^a(t), j_0^b(0)] \rangle$, when current operator commutes with the full Hamiltonian, $[j_0^a(t), j_0^a(0)]$ vanishes and so does any non-trivial optical response. However, when the BCS Hamiltonian is treated within the mean-field approximation, the current operator with supercurrent flowing in the system will not in general commute with the approximate mean field Hamiltonian, leading to a contradiction with the more general principles derived from the full Hamiltonian. This issue, related to gauge invariance, has been recognized in the early days of BCS theory and a solution has been proposed by Nambu \cite{nambuQuasiParticlesGaugeInvariance1960}, restoring gauge invariance in the mean-field BCS formalism by including the vertex correction shown in the main text. Therefore, to obtain physically correct result, we always include this vertex correction in the calculations shown in the main text.

\begin{figure}
    \centering
    \includegraphics[width=0.75\columnwidth]{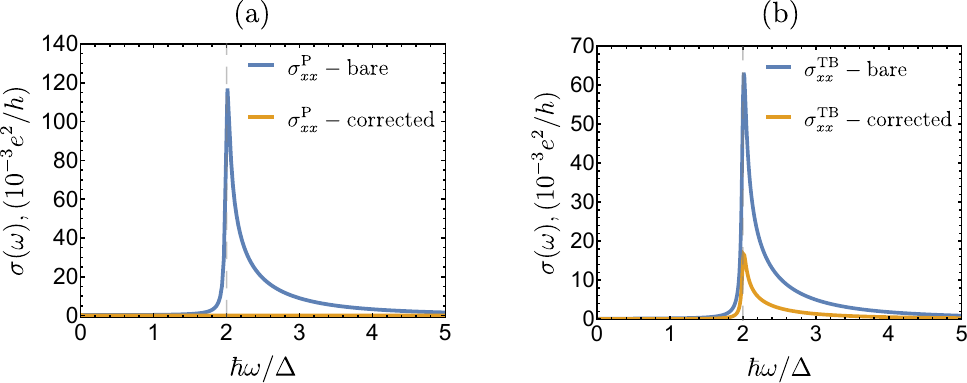}
    \caption{Comparison of optical conductivity with and without vertex correction. (a) System with purely parabolic bands. The optical conductivity calculated using bare current operators is non-zero and given by Eq.~\eqref{eq:SM_sigma_parabolic}. However, when vertex correction is included, the optical conductivity vanishes as expected from Galilean invariance. (b) System with tight-binding band. The optical conductivity calculated with vertex correction is reduced, but doesn't vanish due to deviation from Galilean invariance caused by the crystalline lattice.}
    \label{fig:SM_sigma_comparison}
\end{figure}

Nevertheless, calculations using bare vertices for parabolic systems retain some value due to their analytical tractability, as the functional dependence on frequency and Cooper pair momentum can remain valid for lattice systems that lack Galilean invariance. We start with a single spin-degenerate parabolic band with particles of mass $m$ at chemical potential $\mu$ with a superconducting gap $\Delta$:
\begin{equation}
    H_\text{BdG}^{P}(\mathbf{k}) = \left(\frac{\hbar^2 k^2}{2 m} - \mu \right) \tau_z + \Delta \tau_x
\end{equation}
We can now directly evaluate the Kubo formula as described by the current-current correlation diagram from the main text using bare current operators instead of the one with vertex correction. By computing the momentum integral at $T=0$ we obtain the following results:
\begin{align}
\label{eq:SM_sigma_parabolic}
    \text{Re}\, \sigma_{xx}^\text{P, bare}(\omega) &= \frac{e^2}{h} \frac{\hbar^2 q_x^2}{2m} \frac{8 \pi \Delta^2}{\hbar^2 \omega^2 \sqrt{\hbar^2 \omega^2 - 4 \Delta^2}} \Theta(\omega - 2 \Delta), \\
    \text{Re}\, \sigma_{yy}^\text{P, bare}(\omega) &= 0
\end{align}
This non-zero result is in contradiction with the general consideration based on Galilean invariance. To see how this is changed after introducing vertex correction, in additional to the analysis in the previous section we calculate the corrected expression numerically for the parabolic case as well and show it in Fig.~\ref{fig:SM_sigma_comparison}(a). As expected, the corrected expression vanishes identically in the entire frequency range. However, the frequency dependence given by Eq.~\eqref{eq:SM_sigma_parabolic} still describes the behavior of the system with a tight-binding band as demonstrated in the main text, just with a different numerical prefactor. Similarly, other features like the presence of singularity at the gap edge or vanishing of the response for frequencies within the gap is also retained. As an additional demonstration of the impact of the vertex correction, in Fig.~\ref{fig:SM_sigma_comparison}(b) we show the bare and corrected results of numerical integration for the tight-binding model. The corrected result (the same as presented in the main text in Fig.3(b)) is reduced compared to the bare result, but doesn't vanish like the optical conductivity for the parabolic system. 

We can perform a similar analysis without the vertex correction for the Dirac fermion case, obtaining:
\begin{align}
    \text{Re}\, \sigma_{xx}^\text{D, bare}(\omega) &= \frac{e^2}{h} \frac{3 \pi}{4} \frac{\Delta}{\mu} \frac{\hbar^2 v^2 q_x^2 \Delta}{\hbar^2 \omega^2 \sqrt{\hbar^2 \omega^2 - 4 \Delta^2}} \Theta(\omega - 2 \Delta), \\
    \text{Re}\, \sigma_{yy}^\text{D, bare}(\omega) &= \frac{1}{3} \sigma_{xx}^{D}(\omega)
\end{align}
We see that without the vertex correction, the $\sigma_{xx}$ and $\sigma_{yy}$ components are different from each other. However, while the component parallel to the supercurrent gets reduced by the correction, the component perpendicular to it remains the same. Therefore, after the correction is taken into account, both components of optical conductivity are equal as indicated in the main text.

\begin{figure}
    \centering
    \includegraphics[width=0.4\columnwidth]{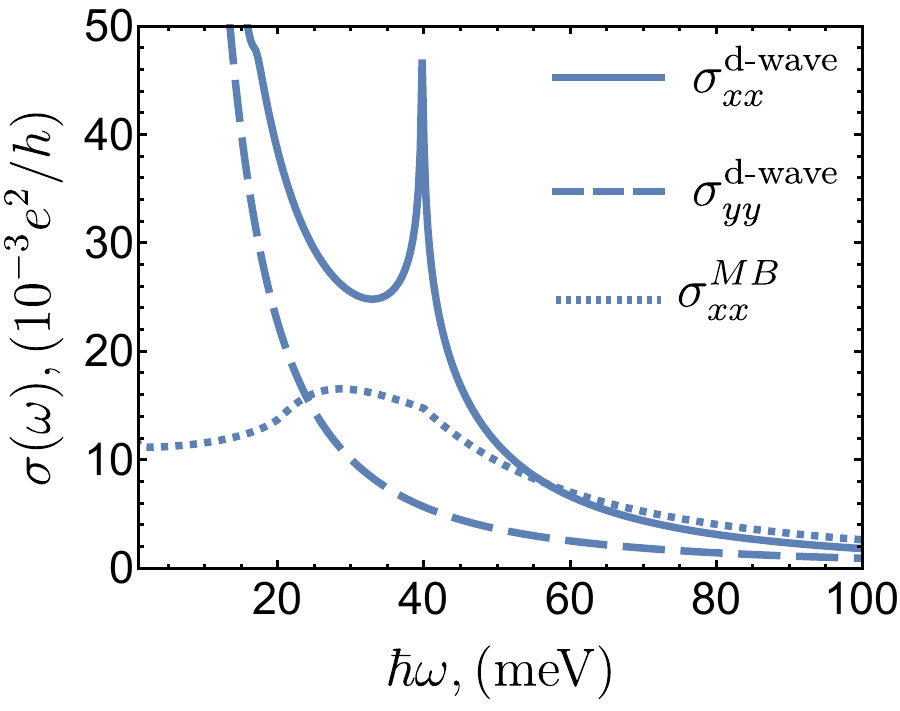}
    \caption{Optical conductivity of d-wave superconductor for parabolic band without the vertex correction. While the presence of vertex correction would cause the contribution to vanish, a Galilean invariance breaking system, such as a simple tight-binding cosine band would result in a similar behavior, with singularity at the gap edge and non-vanishing conductivity inside of the superconducting gap.}
    \label{fig:SM_sigma_dwave}
\end{figure}

Finally, we also consider the case of a spin-degenerate parabolic band with d-wave order parameter:
\begin{equation}
    H_\text{BdG}^\text{d-wave}(\mathbf{k}) = \left(\frac{\hbar^2 k^2}{2 m} - \mu \right) \tau_z + \Delta \frac{k_x^2-k_y^2}{k_F^2} \tau_x
\end{equation}
The parametrization of the order parameter is chosen such that the maximal gap in the spectrum is the same as in the case of both s-wave models, $2 \Delta$. Again, we perform the calculations analytically without the vertex correction, which in this case is more complicated to solve for due to the momentum dependence of the underlying interaction. When discussing the d-wave case based on a standard Kubo formula for non-interacting mean-field system, we can separate the interband and intraband contributions as even a small Cooper pair momentum introduces a Fermi surface due to the nodal nature of the superconducting gap. For the interband contributions we obtain:
\begin{align}
    \text{Re}\, \sigma_{xx,\text{inter}}^\text{d-wave, bare}(\omega) &= \frac{e^2}{h} \frac{\hbar^2 q_x^2}{2m \Delta} \begin{cases} 
      \frac{K\left(\frac{1}{x^2}\right)-E\left(\frac{1}{x^2} \right )}{x}  & x > 1 \\
      \frac{K\left(x^2 \right ) - E\left(x^2 \right )}{x^2} & x < 1 \\
\end{cases}, \\
    \text{Re}\, \sigma_{yy,\text{inter}}^\text{d-wave, bare}(\omega) &= 0
\end{align}
where $K(z)$ and $E(z)$ are complete elliptic integrals of the first and second kind, respectively, and we have introduced symbol $x=\hbar \omega / 2\Delta$ for compactness. Similarly to the parabolic s-wave case, the component perpendicular to the supercurrent vanishes. However, due to the nodal character of the superconducting gap, $\sigma_{xx, \text{inter}}^\text{bare}$ does not vanish immediately for $\hbar \omega < 2\Delta$. Instead, it saturates at a constant value before becoming cut off for $\omega < 2\hbar \sqrt{\mu/m} q_x$ at $T=0$ as both BdG branches are either occupied or empty within that energy window. The behavior at the gap boundary is also singular, but in contrast to s-wave superconductors the singularity is logarithmic $\log(\hbar \omega - 2 \Delta)$. We next calculate the intraband component, which for small Cooper pair momenta is given by:
\begin{equation}
    \text{Re}\, \sigma_{xx,\text{intra}}^\text{d-wave, bare}(\omega) = \text{Re}\, \sigma_{yy,\text{intra}}^\text{d-wave}(\omega) = \frac{e^2}{h} \frac{2 \mu^{3/2} q_x}{\sqrt{m} \Delta} \frac{\eta}{\eta^2 + \omega^2}
\end{equation}
In contrast to the interband contribution, the lowest nonvanishing order of intraband conductivity is linear in $q_x$, compared to $q_x^2$ for all the interband contributions obtained above. The intraband contribution also has a Drude-like behavior, with the peak width determined by the scattering rate $\eta$. As such, it is similar to impurity-induced conductivity of dirty superconductors, and in turn to normal state optical conductivity. Moreover, in this scenario the intraband term is the same to the lowest order in $q_x$ for the components parallel and perpendicular to the supercurrent. The total uncorrected optical conductivity is reached by summing both interband and intraband terms, and is presented in Fig.~\ref{fig:SM_sigma_dwave}. While the results obtained above are without the vertex correction, which would cause the interband contribution to vanish due to Galilean invariance, one can expect a similar behavior for the tight-binding case, in analogy to the s-wave superconductor. Keeping that in mind, both types of contributions can be distinguished by their different dependence on the Cooper pair momentum (and consequently external magnetic field) as well as by the absence of interband contribution for components perpendicular to the current. The exact ratio of these two contributions is also dependent on the purity of the superconductor: for very clean systems the intraband term will be limited to only the lowest frequencies, and the interband term will dominate.

\end{document}